\documentclass{arxivstyle}

\usepackage[OT1]{fontenc} 
\usepackage{graphicx}
\usepackage[dvipsnames]{xcolor}

\title{Quantum Walk Comb in a Fast Gain Laser}
\author [*$\dagger$] {Ina Heckelmann}
\author [$\dagger$] {Mathieu Bertrand}
\author [$\dagger$] {Alexander Dikopoltsev}
\author [ \hspace{-0.7ex}] {Mattias Beck}
\author [ \hspace{-0.7ex}] {\\Giacomo Scalari}
\author [*] {Jérôme Faist}

\affil[ \hspace{-0.7ex}]{Institute of Quantum Electronics, ETH Zürich, 8093 Zürich, Switzerland}
\affil[ \hspace{-0.7ex}]{Quantum Center, ETH Zürich, 8093 Zürich, Switzerland}

\affil[*]{\normalfont corresponding authors: iheckelmann@phys.ethz.ch, jfaist@phys.ethz.ch}
\affil[$\dagger$]{\normalfont These authors contributed equally to this work}

\newcommand{\one}{\textcolor{Gray}}

\begin{document}

\maketitle

\vspace{1pt}
\begin{center}
    \textbf{One Sentence Summary} \\
    A quantum walk in a synthetic space of a ring laser is stabilised by fast gain, \\producing a broadband, stable frequency comb.
\end{center}
\vspace{5pt}

\begin{abstract}
    Synthetic lattices in photonics enable the exploration of light states in new dimensions, transcending phenomena common only to physical space. We propose and demonstrate a quantum walk comb in synthetic frequency space formed by externally modulating a ring-shaped semiconductor laser with ultrafast recovery times. The initially ballistic quantum walk does not dissipate into low supermode states of the synthetic lattice; instead, the state stabilizes in a broad frequency comb, unlocking the full potential of the synthetic frequency lattice. Ou device produces a low-noise, nearly-flat broadband comb (reaching 100 cm\textsuperscript{-1} bandwidth) and offers a promising platform to generate broadband, tunable and stable frequency combs.
\end{abstract}

The random walk is a fundamental concept in mathematics and physics, describing a walker taking random steps in a discretised space. The spread of a classical random walker is characterised by a $\sqrt{N}$ standard deviation after N steps \cite{spitzer_principles_2013}; it can, however, be surpassed by its quantum counterpart, which permits the particle to behave as a wave and interfere with itself \cite{aharonov_quantum_1993}. In this quantum walk, the standard deviation of the position scales linearly with the number of steps, enabling a so-called ballistic spread that facilitates the "quantum speed-up" of search algorithms based on quantum walks \cite{shenvi_quantum_2003}. As coherent light mimics the interference of quantum particles while benefiting from the simpler preparation and manipulation of the light states, photonic lattices provide a natural platform to study quantum walks \cite{schreiber_photons_2010,rai_transport_2008}. This line of research has produced novel phenomena, such as quantum-to-classical walk transitions \cite{eichelkraut_mobility_2013}, quantum walks of correlated photons \cite{peruzzo_quantum_2010}, as well as the experimental study of boson interactions in photonic systems \cite{lahini_quantum_2012}.

In such systems, the study of quantum walks of light was limited to the type of interactions and non-linearities that experiments in real space could provide. However, by coupling accessible degrees of freedom of light, it is possible to construct a synthetic dimension which enables the realisation of previously unattainable physical phenomena. This approach leverages the flexibility and controllability of synthetic dimensions to engineer photonic lattices that effectively overcome the geometric limitations of the system \cite{celi_synthetic_2014,ozawa_synthetic_2016,yuan_photonic_2016,ehrhardt_perspective_2023}. In this way, synthetic dimensions allow for the observation of new physics like parity-time-symmetric lattices \cite{regensburger_paritytime_2012}, complex-valued long-range couplings \cite{bell_spectral_2017}, higher-dimensional photonic crystals \cite{dutt_single_2020,hu_realization_2020}, non-Hermitian funnelling of light \cite{weidemann_topological_2020} and even three-dimensional topological insulators \cite{lustig_photonic_2022}.

\begin{figure}[h!]
    \centering
    \includegraphics[width=1.0\textwidth]{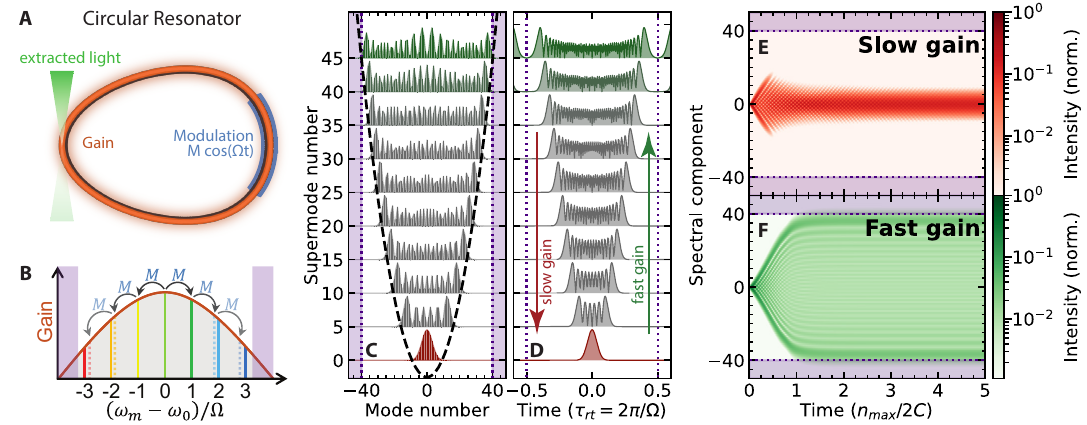}
    \caption{\textbf{Stabilisation of Quantum Walks in synthetic photonic lattices}. The optical modes in a deformed circular cavity (\textbf{A}) are coupled via resonant modulation, inducing nearest-neighbour gain and the proliferation of a synthetic photonic lattice in frequency space, whose bandwidth is fundamentally limited by the dispersion (indigo) (\textbf{B}). In analogy to the quantum harmonic oscillator, the 0\textsuperscript{th} order supermode corresponds to a Gaussian pulse (red) in the discretised frequency (\textbf{C}) and continuous time domain (\textbf{D}), while constant intensity can only be seen for spectrally broad supermodes (green). After the initial ballistic expansion upon the onset of nearest-neighbour coupling \textit{C}, dissipation give rise to a stabilisation of the spectrum. While it collapses to a narrow Gaussian mode for typical slow-gain dissipative systems (\textbf{E}), a superposition of broad supermodes can be observed when a fast gain is employed instead (\textbf{F}).}
    \label{fig1:concept}
\end{figure}

For a synthetic dimension formed by the modes of a resonator, such as frequency ladders in optical ring cavities (Fig. \ref{fig1:concept}A), a phase modulation (induced, for instance, through the electro-optical effect or the laser's pump) can introduce coupling in the chain, giving rise to continuous-time quantum walk dynamics \cite{rai_transport_2008}. For a fixed coupling C, the broadest achievable state is limited by the dispersion in the system, which causes mode resonances to move apart, resulting in less efficient coupling (Fig. \ref{fig1:concept}B) \cite{yuan_photonic_2016}. Qualitatively, dispersion induces an effective potential well in frequency space, while the width of the band created by modulation limits the maximum achievable "kinetic energy" and, consequently, the system's bandwidth. Through this argument, for instance, quadratic dispersion will constrain the highest accessible mode to $N_{max}=2\sqrt{M/D}$, where $M = C/2$ represents the modulation strength and \textit{D} signifies dispersion. This system can be analogously mapped to a quantum harmonic oscillator, which can only occupy a finite number of possible modes (Fig. \ref{fig1:concept}C). 

In harmonic oscillator solutions, only the zeroth order mode is a minimum uncertainty state, being narrowest both in frequency space and reciprocal time, while modes higher on the energy ladder exhibit simultaneously broader spectral and temporal widths (Fig. \ref{fig1:concept}C,D).  Introducing linear non-Hermiticity to the system in the form of frequency-dependent gain and dissipation leads to a contraction of the final state to a narrow Gaussian profile (i.e. a pulse) when reaching the state at the bottom of the dispersion trap. Dynamically, this behaviour is characterised by a transition of the initial ballistic broadening of the quantum walk to a regime of diffusive propagation \cite{eichelkraut_mobility_2013,dikopoltsev_observation_2022} (Fig. \ref{fig1:concept}E). Previously, a similar mapping considering gain curvature instead of dispersion was established \cite{haus_theory_1975}, demonstrating the dissipation to low supermodes in frequency lattices and an M\textsuperscript{1/4}-dependence of the bandwidth \cite{kuizenga_fm_1970} (Fig. S13). The stabilisation in low modes exposes the inherent limitations of such systems in reaching their intrinsic bandwidth.  

\vspace{-1ex}{\subsubsection*{Generating the synthetic frequency lattice}}\vspace{-1ex}

Here, we demonstrate both theoretically and experimentally that the full potential of the synthetic frequency lattice can be unlocked by employing ultrafast saturable gain. The giant third-order nonlinearity of such gain locks the resonator modes and effectively counteracts the dispersion. After the initial single mode performs a spectral quantum walk and expands ballistically, the laser spectrum stabilises on the highest energy supermodes, reaching the predicted fundamental bandwidth limit with a flat spectral envelope and near constant intensity operation (Fig. \ref{fig1:concept}F). This is in contrast with broadband optical frequency combs \cite{udem_optical_2002,diddams_optical_2020,chang_integrated_2022} based on pulsed mode-locked lasers \cite{keller_recent_2003} or on nonlinear generation in microresonators \cite{delhaye_optical_2007,stern_battery-operated_2018}, which typically stabilise into a peaked spectrum typical of a pulsed state. Additionally, in the proposed quantum walk comb, the bandwidth of the resulting low-noise frequency comb is shown to be continuously tunable by varying the radio-frequency (RF) injection.

We consider a synthetic lattice comprised of the resonator modes of an active slightly-deformed circular cavity (Fig. \ref{fig1:concept}A, Fig. S1 and S2), realised using a Quantum Cascade Laser (QCL) emitting in the mid-infrared spectral region. QCLs are semiconductor lasers that provide a flexible platform for high-power combs in the molecular fingerprint region with small device footprints \cite{faist_quantum_1994}. Intriguingly, their gain mechanism is based on intraband transitions and thus differs fundamentally from typical interband lasers, resulting in extremely fast gain recovery times in the order of hundreds of femtoseconds. This type of gain in a Fabry-Perot device will yield frequency modulated (FM) combs \cite{hugi_mid-infrared_2012}, created by four-wave mixing non-linearities arising from spatial hole burning  \cite{khurgin_coherent_2014,gordon_multimode_2008}. To enable linear coupling among lattice sites in the synthetic space and observe the continuous-time quantum walk, we adopted a circular cavity geometry, that suppresses spatial hole burning. Circular QCLs have recently attracted considerable interest since they were shown to host the formation of self-emerging cavity solitons, previously seen in passive microresonators \cite{kippenberg_dissipative_2018}, resulting from the delicate balance between anomalous group velocity dispersion and gain non-linearity \cite{meng_dissipative_2022,kacmoli_unidirectional_2022,piccardo_frequency_2020,micheletti_terahertz_2023}. For our work, the lasers were fabricated using the inverted buried-heterostructure process to achieve smooth sidewalls and hence guarantee unidirectional single-mode lasing \cite{beck_continuous_2002}. In circular cavities where backscattering and hence coupling between counter-propagating modes is sufficiently low, spontaneous symmetry breaking \cite{meng_dissipative_2022} occurs close above threshold, enabling the laser to operate in a unidirectional single-mode state (Fig. S3) \cite{seitner_backscattering-induced_2023}. As a result, it is only when we introduce additional RF modulation at the cavity resonance frequency, that a frequency comb with a broad, highly tuneable, and predictable bandwidth emerges.

When light is extracted from such circular cavities using only bending losses, the output is isotropic and was previously shown to be restricted to the milliwatt level \cite{meng_dissipative_2022}. To vary the cavity circumference, and consequently the roundtrip time, without experiencing an exponential decrease in outcoupled light intensity, we adapted the geometry of a Hügelschäffer egg (Table S1, Fig. S2) \cite{schmidbauer_exakte_1948}. In doing so, the higher curvature at the tip of the eggs creates a designated out-coupling point. Coupling of the resonator modes in the ring QCL was achieved through near-resonant RF current modulation, which provides an efficient phase modulation of the lasing electric field. This effect is enabled by the active region itself, as gain modulation is translated into phase modulation through a non-vanishing linewidth enhancement factor \cite{opacak_frequency_2021,franckie_sensitive_2023}.

\vspace{3ex}{\subsubsection*{Characterising the spectral output}\vspace{-1ex}

To study quantum walk combs, steady-state spectra of our device were measured with Fourier-transform infrared spectroscopy (FTIR) for both a sweep of the injected frequency through the cavity resonance at 15.772\,GHz, and without any applied modulation. While the free-running device operates as a single-mode laser at 1280\,cm$^{-1}$ at the investigated working point at 1.456\,A and -16\,°C (Fig. S3C), it gradually and reliably broadens to a bandwidth of 70\,cm$^{-1}$ when RF modulation is enabled and tuned to resonance (Fig. \ref{fig4:steady-state}B). For resonant injection (working point 2), the spectral envelope showcases lobes on both sides and a nearly flat centre, as expected for a quantum walk stabilizing on a highly excited state \cite{aharonov_quantum_1993,spitzer_principles_2013}. For this resonant injection, a participation ratio, given by $\left( \sum_m I_m \right)^2 / \sum_m I_m^2$, where $I_m$ is the intensity of the m-th optical mode, of 74.3 is computed for the 132 involved lattice sides, indicating a highly delocalised, i.e. flat, spectrum. As we detune from resonance, the spectra narrow continuously and display an imbalance (working points 3,4), which rectifies far from resonance (working point 1). 

\begin{figure}[h!]
    \centering
    \includegraphics[width=1\textwidth]{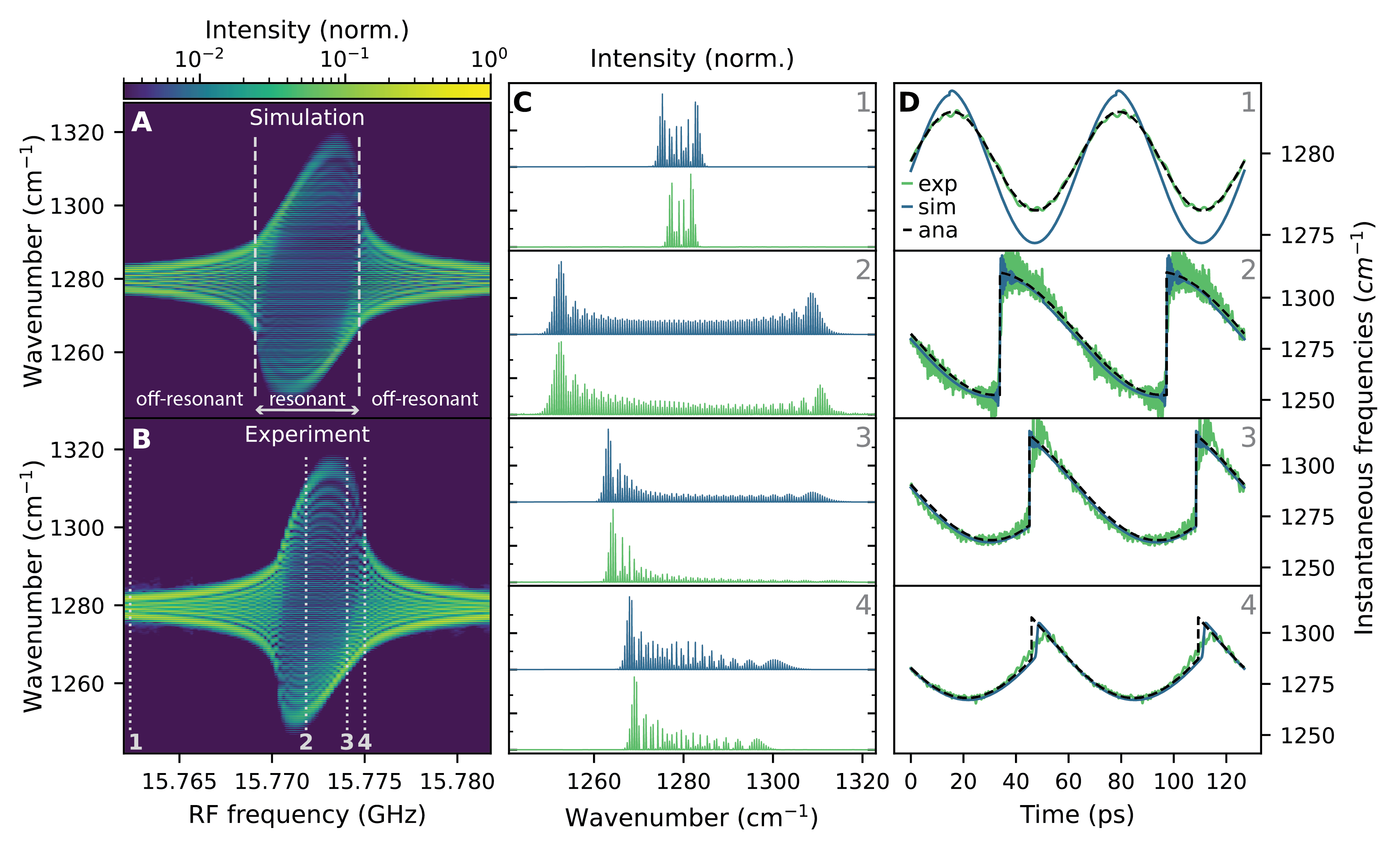}
    \caption{\textbf{Steady-state of the Quantum Walk Comb}. Spectra for sweeps of the injection frequency, as predicted by simulations based on the Maxwell-Bloch equations (\textbf{A}) and measured in FTIR (\textbf{B}) with distinct on- and off-resonant lasing regimes. Spectra (\textbf{C}, linear scale) and instantaneous frequencies (\textbf{D}) at the indicated working points (vertical lines in \textbf{B}) display narrow Bessel-like spectra and sinusoidal phases in the off-resonant regime (1) and discontinuous phases at half the roundtrip frequency in the novel on-resonance regime (2,3). The transition between the two lasing states occurs at a detuning of 2.8\,MHz (4) with unbalanced spectra and an instantaneous frequency with negligible jumps. The experimental instantaneous frequencies ('exp', green) are well-described by both numerical simulations ('sim', blue) and the derived analytical expression ('ana', black).}
    \label{fig4:steady-state}
\end{figure} 

To theoretically investigate the comb formation, we modelled the propagating field in the dispersive waveguide of the active ring QCL based on \cite{burghoff_unraveling_2020,opacak_theory_2019}, while neglecting spatial hole burning due to the absence of any counterpropagating wave. Here, the phase of the field follows the modulation with depth \textit{M} and detuning $\Delta \Omega$ from the cavity resonance. Through this model, assuming instantaneous gain for QCLs and thus a constant intensity, we calculated the steady state of the electric field in the co-rotating frame of reference (derivation in \cite{see_SM}). By adjusting only a small set of parameters, we accurately reproduced the experimental spectra (Fig. \ref{fig4:steady-state}A,C). The bandwidth of the comb under resonant injection reaches the previously discussed fundamental limit of the synthetic lattice.

Upon closer examination of the phases predicted by our model, we observe the existence of two distinct lasing regimes, which depend on the detuning $\Delta \Omega$ of the injected frequency from the cavity resonance (Fig. S11). In the case of highly off-resonant injection, the comb state exhibits Bessel-like spectra and sinusoidal instantaneous frequency with the periodicity of the injected signal, akin to the characteristics previously observed in pure FM lasers \cite{harris_theory_1965}. However, within a range of near-resonant injections, we discovered a novel FM regime characterized by a discontinuous cosine phase at only half the modulation frequency  (Fig. \ref{fig4:steady-state}D). The phase adheres to the following analytically derived relation ('ana' in Fig. \ref{fig4:steady-state}D), assuming a constant intensity: 

\begin{equation}
    \Phi (z, t) = N_1 \cdot \left( \frac{Kz - \Delta \Omega t}{2} \right) + N_2 \cdot \cos\left( \frac{Kz - \Delta \Omega t}{2} \right)
\end{equation}

where \textit{K} represents the cavity wavenumber, and $N_1$ and $N_2$ denote the amplitudes of phase modulation, which directly rely on the modulation depth \textit{M} and dispersion $\beta$  (derivation in \cite{see_SM}).

To experimentally corroborate the presence of these separate lasing regimes, we conducted phase measurements of the laser using shifted-wave interference Fourier-transform spectroscopy (SWIFTS) \cite{burghoff_evaluating_2015} at various injection frequencies around the resonance ('exp' in Fig. \ref{fig4:steady-state}D, Fig. S4). The measured instantaneous frequencies align remarkably well with the numerically simulated and analytically derived solutions for both lasing regimes. Furthermore, coherence was confirmed through these measurements, encompassing the entire spectral bandwidth (Fig. S6). However, it's important to note that our experiment measures the collective behaviour of the modes, resembling that of a single particle, and does not detect the presence of non-classical correlations. Our steady-state measurements of the frequency comb convincingly confirmed its coherence, continuous and predictable tuning, wide bandwidth, and excellent agreement with our theoretical models.

\vspace{-1ex}{\subsubsection*{Resolving the Photonic Quantum Walk}\vspace{-1ex}}

Investigating the role of quantum walk dynamics in shaping the steady-state spectra, we conducted measurements and simulations to explore the underlying transient processes. To track the dynamical comb broadening under RF injection, we employed boxcar averaging in conjunction with an FTIR at 10\,ns temporal resolution (Fig. S7, S8). The experimental spectra (Fig. \ref{fig2:QW}C,D) exhibit a ballistic expansion to the full 70\,cm$^{-1}$ bandwidth within the initial 250\,ns, followed by a subsequent locking that maintains this spectral state for the investigated times.  

\begin{figure}[h!]
    \centering
    \includegraphics[width=0.62\textwidth]{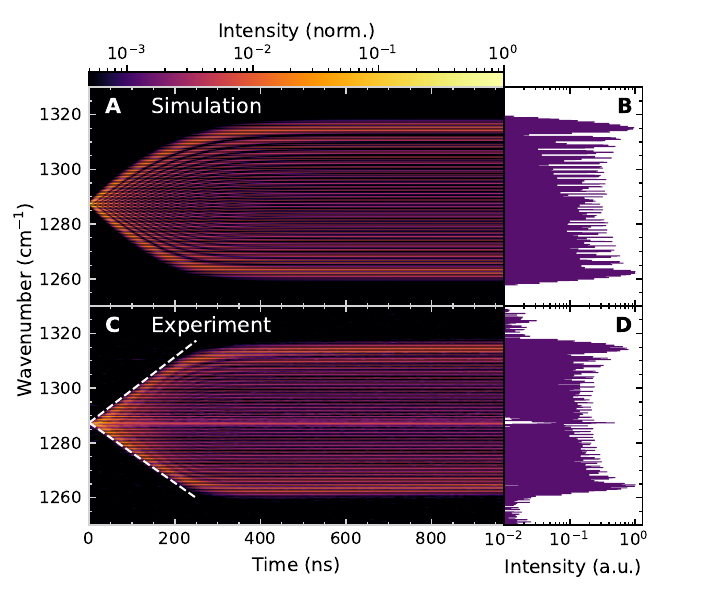}
    \caption{\textbf{Quantum Walk and Locking}. Simulations based on the Maxwell-Bloch master equation (\textbf{A}) reproduce the ballistic expansion and subsequent locking to a spectrally broad supermode (\textbf{B}) that is observed in the time-resolved spectral measurements (\textbf{C},\textbf{D}). The dashed white lines indicate the ballistic broadening of the Quantum Walk.}
    \label{fig2:QW}
\end{figure} 

Building upon these experimental findings, we now delve into the underlying theoretical model, using the Maxwell-Bloch framework to unravel the dynamics of the quantum walk responsible for the observed spectral broadening during RF modulation. As mentioned earlier, the free-running laser solely occupies the resonator mode with the highest gain, but when resonant injection is introduced, mixing of the optical photons and modulation quanta $\Omega$ leads to nearest-neighbour gain in the photonic lattice. Through discrete hopping between resonator modes, the photons perform a continuous-time quantum walk, resulting in a ballistic broadening of the spectrum, whose bandwidth increases linearly with the number of quantum walk steps. Once the spectrum reaches the frequency boundary dictated by the dispersion and modulation parameters, it stabilises into a broad supermode both spectrally and temporally, maintaining this state for subsequent time intervals (Fig. \ref{fig2:QW}A,B). The only clear difference between the experiment and our model is the much slower experimental decay of the central mode at 1287\,cm$^{-1}$ in the measured spectra. The otherwise remarkable agreement between our simulated and experimental spectra provides compelling evidence for the rapid expansion and stable locking of the quantum walk comb. 

\vspace{-1ex}{\subsubsection*{Tunability and stability}\vspace{-1ex}}

Solving the Maxwell-Bloch equations at resonance, we can determine the state on which the quantum walk stabilizes, leading to a bandwidth dependence of $\sqrt{M/D}$ (derivation in \cite{see_SM}). This power dependence of the bandwidth offers a quadratic improvement over the M\textsuperscript{1/4}-relation typically seen in active mode-locking of lasers \cite{kuizenga_fm_1970}. Since the modulation depth \textit{M} is proportional to the square root of the injected power, a parameter that we experimentally control, we anticipate observing a $P^{1/4}$-dependence of the bandwidth. Confirming this expectation, a power sweep at resonance validated the predicted scaling behaviour, with saturation occurring around 66\,cm$^{-1}$ at 31.5\,dBm for the investigated device (Fig. \ref{fig:bandwidth}). We attribute the deviations from the expected behaviour at high power to non-linearities in the modulation dynamics of the QCL active region, as well as higher-order dispersion. The simple relationship between modulation power and optical bandwidth allows the power per comb line to be controlled flexibly and with very short time delays, since the average laser power remains approximately constant. On the same figure, we report the results obtained on a device which exhibited an even broader bandwidth that reached 100 cm$^{-1}$; however, it could not be driven to full resonance. The most likely reason is a larger amount of third-order dispersion resulting from the wider waveguide. 

\begin{figure}[h!]
    \centering
    \includegraphics[width=0.67\textwidth]{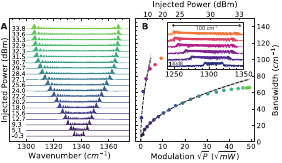}
    \caption{\textbf{Quantum walk comb device under RF injection}. The progressive broadening of the spectra with increasing injection power (\textbf{A}, linear scale) follows the predicted $M^{1/2}$ (and $P^{1/4}$) dependence up until saturation around 31.5\,dBm (\textbf{B}), lower curve. A second device showed broadening up to 100 cm$^{-1}$ but could not be driven at resonance, as shown by the typical asymmetric spectra (inset, logarithmic scale).}
    \label{fig:bandwidth}
\end{figure} 

Under resonant injection, the comb demonstrates the same level of stability as the free-running single-mode laser, providing evidence that the modulation itself does not introduce additional noise contributions. In particular, the measured amplitude noise power spectral density is unchanged by modulation, is shot-noise limited above approximately 100\,kHz and exhibits the distinct 1/f-dependence for lower frequencies (Fig. S9). These remarkable characteristics attest to the exceptional stability of the comb operation. With the inherent stability of the RF modulation faithfully transferred to the optical spectrum, the quantum walk comb presents a highly promising platform for the generation of broadband, tunable, and stable frequency combs.

\vspace{-2ex}{\subsubsection*{Concluding remarks}}\vspace{-1ex}

We have observed in real-time the ballistic expansion and stabilization of the spectrum in a synthetic photonic lattice operating in a laser upon resonant phase modulation, a display that can be ascribed to nonlinear continuous-time quantum walks. In particular, we demonstrate that in our ring QCL structure, the fast gain saturation that constrains the output to a near-constant intensity leads the quantum walk to the spectrally broadest state allowed by dispersion. The quantum walk comb exhibits high predictability, tunability, and stability. In this way, we address an issue raised many years ago: active mode-locked devices with slow gain recovery operate on the fundamental narrowest Gaussian mode with limited comb bandwidth, all other modes being unstable \cite{haus_theory_1975}. Leveraging the insights gained from our work, deliberate design strategies for wide-bandwidth comb sources could open up applications in fields such as spectroscopy \cite{picque_frequency_2019} and telecommunications \cite{pfeifle_coherent_2014}, as the principles of this work are not limited to QCLs but could equally exploit fast gain components existing in interband devices to expand to other wavelength regimes \cite{senica_frequency-modulated_2023}. Furthermore,  this approach holds promise for the application of quantum walk algorithms in systems featuring designated nonlinear interactions among qubits and can be generalized to higher dimensional lattices in either real space with coupled resonators or synthetic space using additional modulation frequencies\cite{hu_realization_2020}.

\clearpage

\bibliographystyle{bibstyle}

\clearpage

\subsubsection*{Funding}

MIRAQLS: Staatssekretariat für Bildung, Forschung und Innovation SBFI: 22.00182 \\(in collaboration with EU - Grant Agreement: 101070700)

Swiss National Science Foundation: 212735

Innosuisse: Innovation Project 52899.1 IP-ENG, Agreement Number 2155008433 "High yield QCL Combs"

ETH Fellowship program: 22-1 FEL-46 (AD)

\subsubsection*{Author Acknowledgements}

IH processed the devices, performed the characterisations and wrote the original draft. MBer performed and analysed the SWIFTS and time-resolved measurements with assistance from IH. AD provided predictive models and performed the simulations. IH, MBer, AD and MBec wrote the Supplementary Materials. MBec grew the QCL wafer and consulted in the fabrication. AD, GS and JF acquired the funding, administrated and supervised the project. JF conceptualised the idea. All authors contributed to the interpretation of the results and the review and editing of the draft.

\subsubsection*{Competing interests}

The authors declare that they have no competing interests.

\subsubsection*{Data and materials availability}

Data that support the findings of this article and codes are available in the ETH Research Collection \cite{heckelmann_eth_nodate}.

\end{document}